\documentclass[aps,prl,reprint,amsmath,amssymb,superscriptaddress]{revtex4-1}

\usepackage{graphicx}
\usepackage[pdftex,dvipsnames]{xcolor}
\usepackage{hyperref}

\begin{document}

\title{Zero-Energy Modes from Coalescing Andreev States \\ in a Two-Dimensional Semiconductor-Superconductor Hybrid Platform}

\author{H.~J.~Suominen}
\author{M. Kjaergaard}\thanks{Now at Department of Physics, Massachusetts Institute of Technology, Cambridge, MA 02139, USA}
\affiliation{Center for Quantum Devices and Station Q Copenhagen, Niels Bohr Institute, University of Copenhagen, Universitetsparken 5, 2100 Copenhagen, Denmark}

\author{A.~R.~Hamilton}
\affiliation{Center for Quantum Devices and Station Q Copenhagen, Niels Bohr Institute, University of Copenhagen, Universitetsparken 5, 2100 Copenhagen, Denmark}
\affiliation{School of Physics, University of New South Wales, Sydney, New South Wales 2052, Australia}

\author{J.~Shabani}
\affiliation{California NanoSystems Institute, University of California, Santa Barbara, California 93106, USA}
\affiliation{Center for Quantum Phenomena, Physics Department, New York University, New York 10031, USA}

\author{C.~J.~Palmstr\o{}m}
\affiliation{California NanoSystems Institute, University of California, Santa Barbara, California 93106, USA}
\affiliation{Department of Electrical Engineering, University of California, Santa Barbara, California 93106, USA}
\affiliation{Materials Department, University of California, Santa Barbara, California 93106, USA}

\author{C.~M.~Marcus}
\author{F.~Nichele}
\email{fnichele@nbi.ku.dk}
\affiliation{Center for Quantum Devices and Station Q Copenhagen, Niels Bohr Institute, University of Copenhagen, Universitetsparken 5, 2100 Copenhagen, Denmark}

\date{\today}

\begin{abstract}
We investigate zero-bias conductance peaks that arise from coalescing subgap Andreev states, consistent with emerging Majorana zero modes, in hybrid semiconductor-superconductor wires defined in a two-dimensional InAs/Al heterostructure using top-down lithography and gating. The measurements indicate a hard superconducting gap, ballistic tunneling contact, and in-plane critical fields up to $3$~T. Top-down lithography allows complex geometries, branched structures, and straightforward scaling to multicomponent devices compared to structures made from assembled nanowires.
\end{abstract}
\maketitle

There is growing interest in material systems that both support Majorana zero modes (MZMs) relevant for topological quantum computing \cite{Kitaev2003}, and can be fabricated to provide branched, complex, and scalable geometries. Emerging as zero-energy states in one-dimensional semiconductors with induced superconductivity, Zeeman coupling, and spin-orbit interaction \cite{Oreg2010,Lutchyn2010}, MZMs have been tentatively identified in individual InSb or InAs nanowires \cite{Mourik2012,das2012,Deng2012,Churchill2013,Zhang2016}, including recently realized epitaxial hybrids \cite{krogstrup2015,chang2015,Deng2016}. Future tests of non-Abelian statistics will likely involve braiding \cite{alicea2011, Aasen2016} or interferometric measurement \cite{Vijay2016,Karzig2016,Plugge2017}, requiring branched or looped geometries, challenging to realize using individual nanowires or mechanically assembled nanowires networks.

In this Letter, we investigate wirelike devices lithographically defined on a two-dimensional (2D) epitaxial InAs/Al heterostructure \cite{Shabani2015}, a material system that recently yielded devices with highly transparent superconductor-semiconductor interfaces, as demonstrated by near-unity Andreev reflection probability \cite{Kjaergaard2016,Kjaergaard2016b}. In a large magnetic field, a zero-bias peak (ZBP) emerges from coalescing Andreev bound states, consistent with the appearance of MZMs. The ZBP shows stability in gate voltage and magnetic field, distinguishing it from simple zero-crossing Andreev bound states. 

\begin{figure}[ht]
	\includegraphics[]{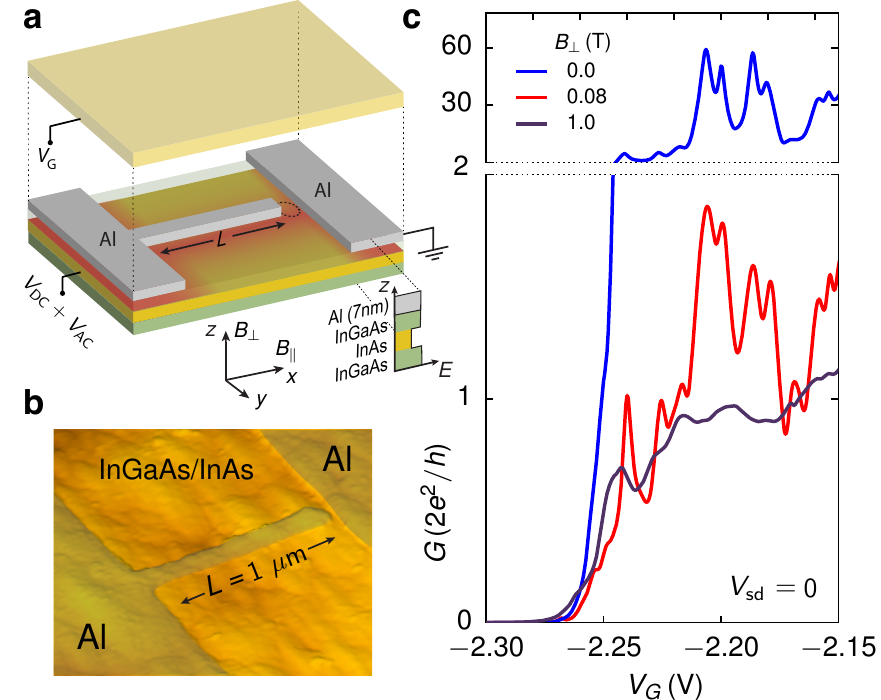}
	\caption{(a) Device schematic indicating the aluminum leads (gray), InAs 2DEG (yellow), InGaAs barrier (green) and top gate (orange). The insulating layer between the device structure and the electrostatic gate has been omitted for clarity. The tunneling probe location is indicated by the dashed circle. Inset: Band alignment as a function of depth $z$ highlighting the finite confining barrier between the Al and InAs. (b) False colored atomic force micrograph of a lithographically identical device before oxide and gate deposition. (c) Conductance as a function of gate voltage for $B=0$ (blue), $B_\perp=0.08~\rm{T}$ (red) and $B_\perp=1~\rm{T}$ (purple).}
	\label{fig:sqpcs}
\end{figure}

\begin{figure*}[t]
 \includegraphics[]{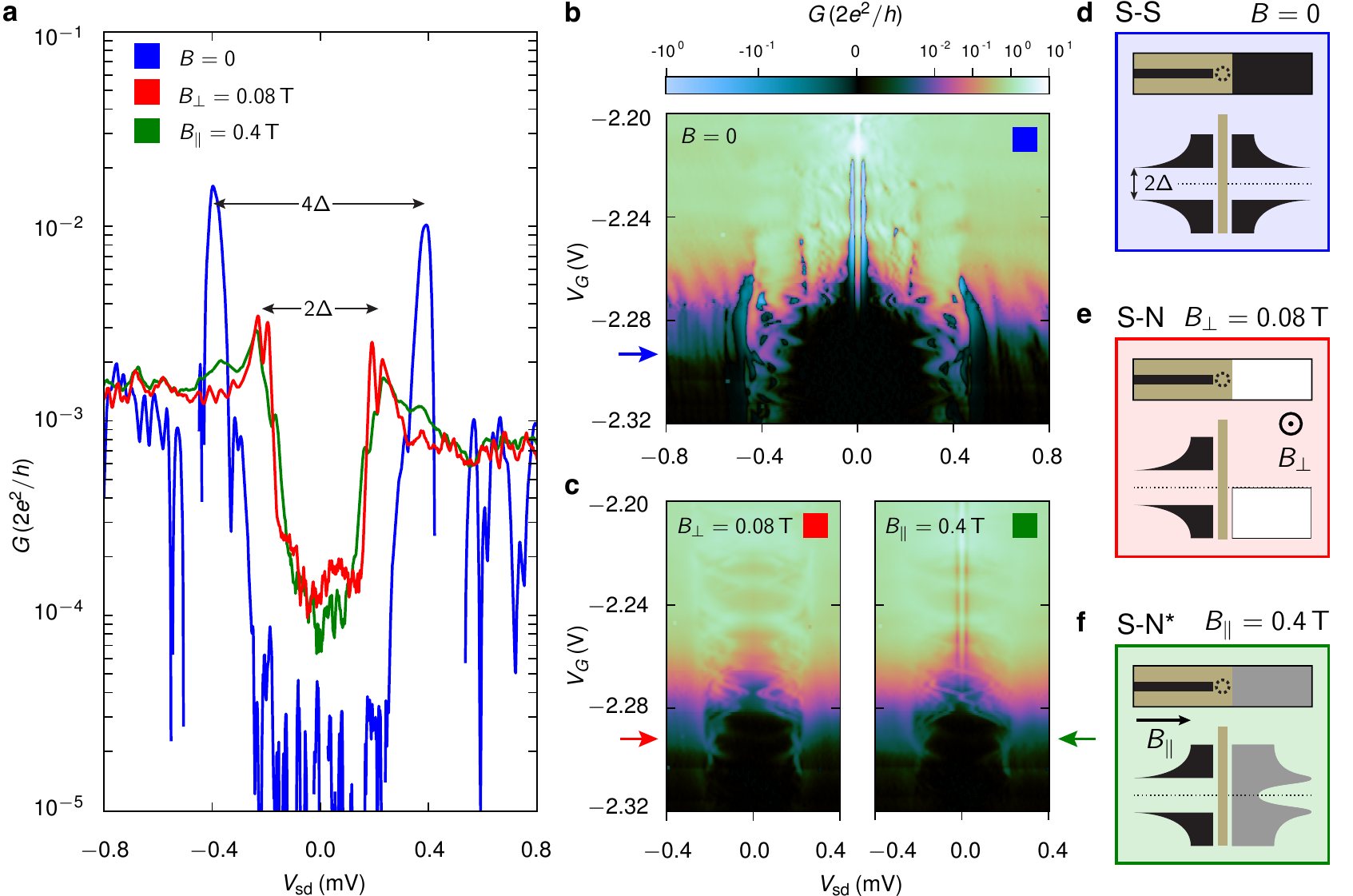}
 \caption{(a) Tunneling spectroscopy of the wire for $B=0$ (blue), $B_\perp = 0.08~\rm{T}$ (red), and $B_\parallel = 0.4~\rm{T}$ aligned along the wire (green). (b),(c) Tunneling spectroscopy of the superconducting gap for the three field configurations in (a). The colors from (a) identify each panel, with arrows indicating the gate voltage location of the traces in (a). A nonlinear color scale is used.
 (d)--(f) Schematic representations of the three regimes of operation shown in (a)--(c) with relative DOS in the wire (left) and Al plane (right). Superconducting Al is represented in black, white indicates that the Al has been driven normal, and gray that the Al is still superconducting but the induced gap is soft.}
 \label{fig:hardness}
\end{figure*}

A schematic of one of the samples is shown in Fig.~\ref{fig:sqpcs}(a), with the heterostructure layers in the inset. The InAs/InGaAs quantum well is close to the surface and covered by a thin layer of epitaxial Al. Large mesas are first etched to isolate individual devices (not shown), then the Al top layer is selectively etched into an effective wire of width $W\sim100~\rm{nm}$ and length $L\sim1~\rm{\mu m}$ [Fig.~\ref{fig:sqpcs}(b)]. One end of the wire is connected to a large Al plane serving as measurement ground. On the other end, an $\sim40~\rm{nm}$ gap [indicated by the dashed circle in Fig.~\ref{fig:sqpcs}(a)] separates the Al wire from the opposing Al plane, acting as voltage source. A global insulating layer and a metallic top gate were then deposited on the entire sample. More details on the sample fabrication are provided in the Supplemental Material \cite{Supplement}.
Initially, the Al wire is surrounded by conductive two-dimensional electron gas (2DEG). Applying a negative potential $V_{G}$ to the top gate, the wide exposed 2DEG regions adjacent to the Al strip are depleted, leaving a narrow conducting InAs channel strongly coupled to the Al. Due to screening by the surrounding Al, conduction through the constriction persists to more negative gate voltages than the 2DEG planes, resulting in a gate voltage range where the wire and Al plane are tunnel coupled. As we will show in the following, the constriction is single mode and quasiballistic. Furthermore, the asymmetric Al regions allow for a useful (and, to our knowledge, novel) magnetic field tuning of the device properties. As the Al strip width $W$ is significantly shorter than the superconducting coherence length $\xi_{\rm Al}\sim 1.6~\rm{\mu m}$ \cite{Merservey1969}, its critical field is enhanced with respect to the Al plane \cite{Moshchalkov1995,Poza1998}. It is then possible by changing the magnetic field strength and orientation to tune the wire-plane configuration from superconductor-superconductor ($S-S$), to superconductor-normal ($S-N$), to normal-normal ($N-N$). We give evidence of this tuning both in the open regime [Fig.~\ref{fig:sqpcs}(c)] and in the tunneling regime [Fig.~\ref{fig:hardness}(a)].

\begin{figure*}[t]
 \includegraphics[]{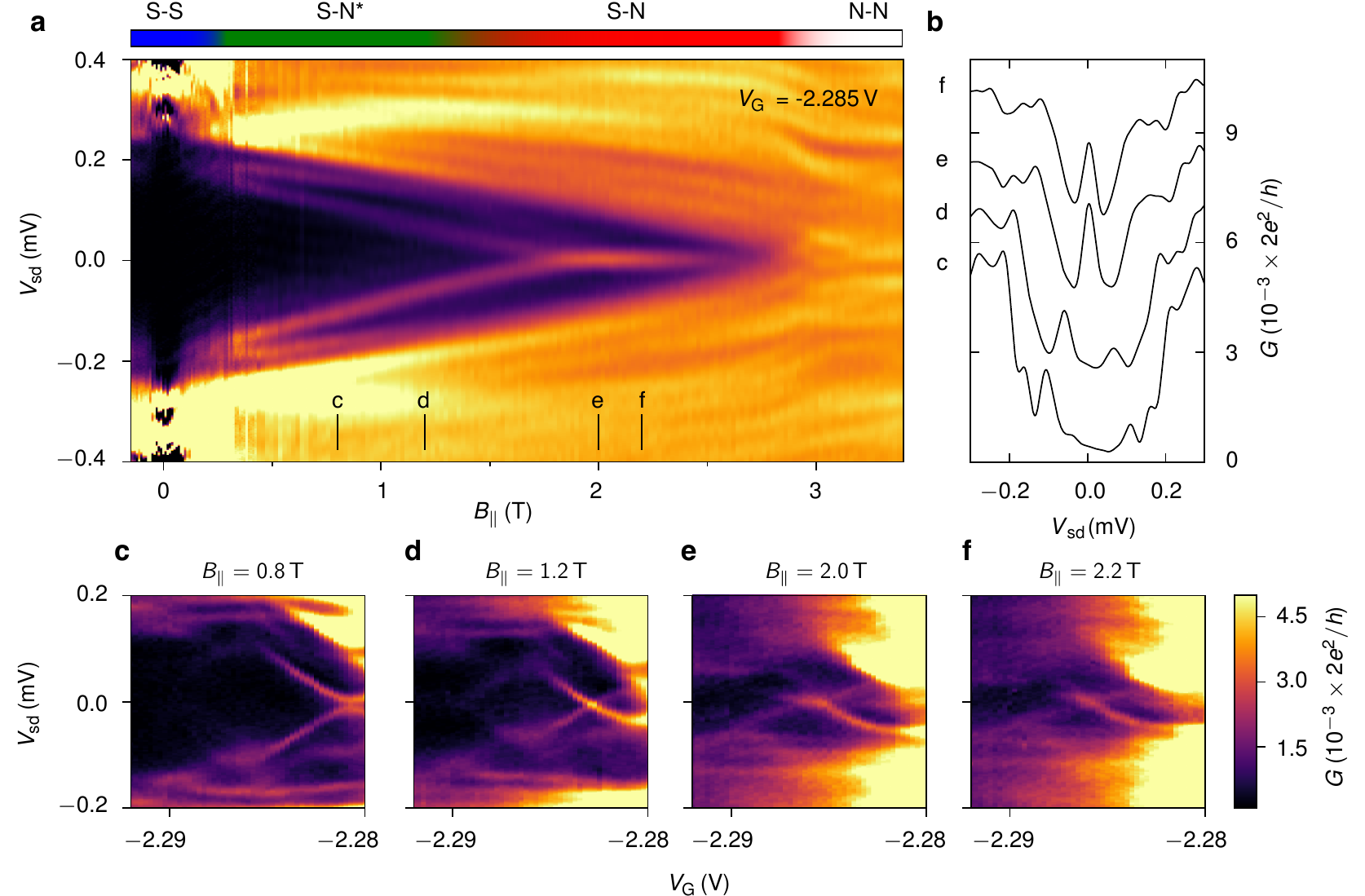}
 \caption{(a) Conductance as a function of source-drain bias and parallel magnetic field. The upper color bar schematically indicates, with reference to Fig.~\ref{fig:hardness}, the DOS configuration in the wire and under the 2D plane. The color scale used is shared with (c)--(f). (b) Line cuts taken at the points indicated in (a). Curves are successively offset by $2.5\times10^{-3}~2e^2/h$. (c)--(f) Stability scans as a function of bias and gate voltage at the field positions indicated in (a).}
 \label{fig:magneticfield}
\end{figure*}

The four-terminal differential conductance of the device as a function of gate voltage is shown in Fig.~\ref{fig:sqpcs}(c). We are interested in the regime close to pinch-off, where the narrow junction is well defined. Applying an out-of-plane field $B_\perp=1~\rm{T}$, superconductivity in the whole system is suppressed, resulting in the $N-N$ configuration. Similar to a conventional quantum point contact, the conductance shows a plateau at $G\sim2e^2/h$, demonstrating the junction probing the wire is single mode and ballistic. In the same gate voltage range, the zero field data ($S-S$ configuration, blue line) show a conductance increase up to $120~e^2/h$, reminiscent of a supercurrent. Finally, setting $B_\perp$ to $0.08~\rm{T}$, the Al plane is driven normal ($B_{\perp,c}\sim0.06~\rm{T}$) while the wire persists in the superconducting regime, resulting in the $S-N$ configuration (red curve). In the S-N configuration the conductance plateau approaches $4e^2/h$, as expected in a single-mode $S-N$ junction with high probability of Andreev reflection \cite{Beenakker1992}, and recently reported in a similar system \cite{Kjaergaard2016}. Andreev reflection has a non-linear dependence on transmission \cite{Beenakker1992}, resulting in the amplification of the conductance resonances in the $S-S$ and $S-N$ case with respect to the $N-N$ case. The resonances visible only for large transmission, are suppressed for source-drain biases larger than the superconducting gap.

The magnetic tuning of the junction is also evident in the tunneling spectroscopy data shown in Fig.~\ref{fig:hardness}(a). At zero field, tunneling conductance in the $S-S$ geometry (blue line) shows a gap of $4\Delta$, where $\Delta=180~\rm{\mu eV}$ is the superconducting gap of the wire and the lead, owing to convolution of two superconducting densities of states, as shown in Fig.~\ref{fig:hardness}(d) \cite{Wolf2011}. In the $S-N$ configuration (red line) the constant density of states in the plane, as shown in Fig.~\ref{fig:hardness}(e), results in a direct measurement of the superconducting gap of the wire. The full gate voltage evolution in the $S-S$ and $S-N$ scenarios is presented in Figs.~\ref{fig:hardness}(b) and \ref{fig:hardness}(c) (left panel), identified by the colored boxes. In both cases, a sharp transition from $G\sim2e^2/h$ to $G\sim0$ is observed at large bias, indicating a clean junction. The $S-S$ configuration also shows for $V_{\rm{sd}} = 0$ and $V_{g}>-2.28~\rm{V}$, a large conductance peak surrounded by regions of negative differential conductance, which is identified as a supercurrent precursor \cite{Iansiti1989}. Similarly, regular subgap features in the open $S-S$ regime are assigned to multiple Andreev reflections. Supercurrent and multiple Andreev reflections disappear in the $S-N$ configuration [Fig.~\ref{fig:hardness}(c), left panel]. 

A particularly interesting situation is obtained for an in-plane field $B_{\parallel}=0.4~\rm{T}$ aligned along the wire, well below the critical field of the large Al plane ($B_{\parallel,c}\sim1.3~\rm{T}$). Tunneling spectroscopy in this regime reveals a $2\Delta$ gap [green line in Fig.~\ref{fig:hardness}(a)] very similar to the $S-N$ configuration discussed previously. On the other hand, conductance in the open regime shows a supercurrent peak [Fig.~\ref{fig:hardness}(c), right panel], a hallmark of the $S-S$ configuration. This seemingly contradictory scenario is readily explained with a superconducting density of states in the large Al regions developing a soft gap in an in-plane field, as shown in Fig.~\ref{fig:hardness}(f). In this configuration referred to as $S-N^\star$, the 2D plane stays superconducting, but in the tunneling regime, it acts as a quasiconstant DOS probing the wire. Independent measurements of the field-induced gap softening in a variety of samples are presented in the Supplemental Material \cite{Supplement}.

We now focus on probing the wire under conditions relevant for topological superconductivity. To enter the topological phase, the theoretical recipe calls for applying a magnetic field perpendicular to the spin-orbit direction. For a Rashba-dominated system, the spin-orbit field is expected to be in the plane of the 2DEG and perpendicular to current flow. We, thus, orient $B_{\parallel}$ along the wire direction using a vector magnet. The topological transition is expected at a field $B_{T}^*=2\sqrt{{\Delta}^2+\mu^2}/g\mu_B$ \cite{Oreg2010}, with $\mu$ the chemical potential, $g$ the $g$-factor of the states in the wire, and $\mu_B$ the Bohr magneton.

Figure~\ref{fig:magneticfield}(a) shows the wire tunneling conductance as a function $B_\parallel$ for a top gate voltage $V_{G} = -2.285~\rm{V}$, setting the constriction in the tunneling regime. The $4\Delta$ transport gap observed for $B_\parallel=0$ collapses to $2\Delta$ by $B_\parallel=0.3~\rm{T}$, attributed to the gap softening under the 2D plane [corresponding to the transition from Figs.~\ref{fig:hardness}(d) to \ref{fig:hardness}(f)]. The 2D plane evolves continuously from a softened gap ($S-N^\star$) into the normal state ($S-N$) by $B_\parallel\sim1.5~\rm{T}$. For $B_\parallel\geq2.9~\rm{T}$, superconductivity in the Al wire is quenched, yielding the $N-N$ state.

Starting from $B_\parallel = 0.4~\rm{T}$ two states emerge from the gap edge and linearly approach $V_{\rm{sd}} = 0$ with an effective $g$-factor $|g^*| = 2\delta V_{\rm{sd}}/\mu_B \delta B =4.2$. At $B_\parallel = 1.8~\rm{T}$, the states merge at zero energy and stick there until the overall gap collapses at $B_\parallel = 2.9~\rm{T}$. Figure~\ref{fig:magneticfield}(b) shows line cuts from Fig.~\ref{fig:magneticfield}(a) at the marked positions.
The two states are symmetrically positioned around zero bias, as expected by particle-hole symmetry, but have different amplitudes due to device asymmetries. Exchanging source and drain contacts reverses the asymmetry. Similar to previous results in nanowires \cite{Deng2016}, the $g$-factor associated to the Majorana precursors is significantly reduced from that of the bulk semiconductor ($g\sim-12$ for InAs). We attribute the reduced $g$-factor as reflecting the electron wave function being partially in the Al and partially in the InAs. 

\begin{figure}[t]
 \includegraphics[]{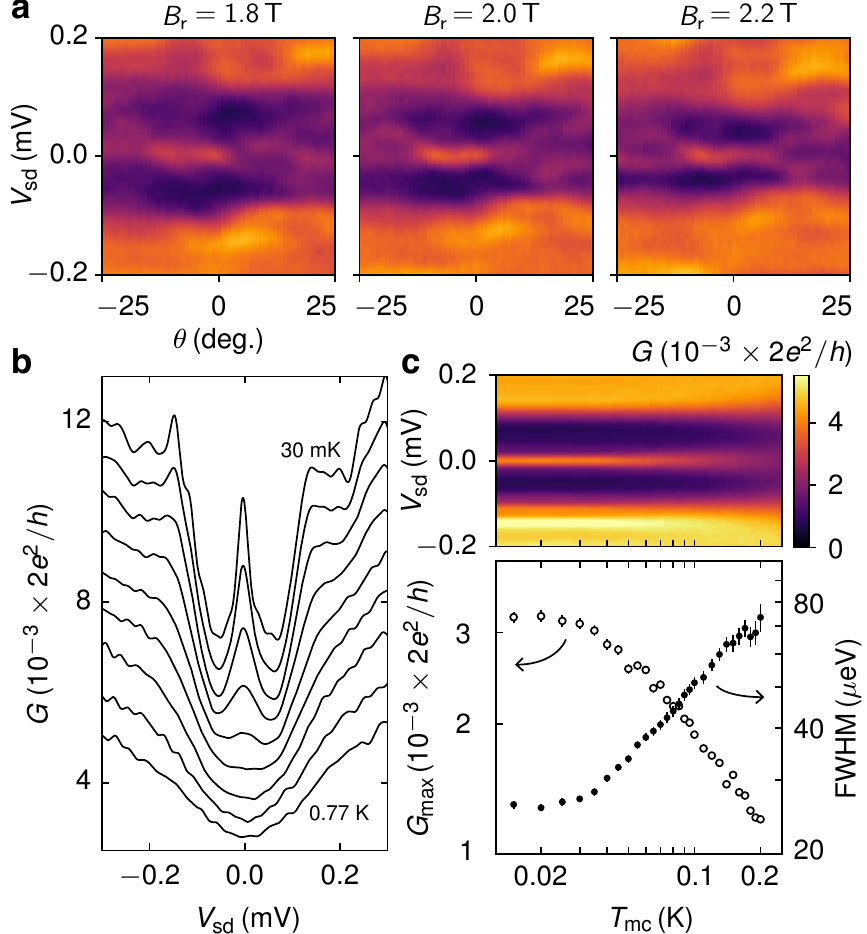}
 \caption{(a) Conductance as a function of bias and in-plane magnetic field orientation $\theta$ for fixed field magnitudes $B_{r}$. $\theta=0$ indicates a field alignment parallel to the wire. (b) Conductance line cuts as a function of bias for fixed values of the mixing chamber temperature $T_{\rm{mc}}$. With decreasing temperature, the ZBP gets sharper and higher. Curves are offset for clarity. (c) Detailed temperature evolution of the ZBP (upper panel), and extracted ZBP height $G_{\rm max}$ and FWHM as a function of $T_{\rm{mc}}$ (lower panel). Note that the vertical axes have logarithmic scales.}
 \label{fig:stability}
\end{figure}

A metric of the ZBP stability was proposed in Ref.~\onlinecite{Lee2014} as the ratio $\eta$ between $g\mu_{B}\Delta B$, the expected Zeeman energy splitting of the Andreev states, and the peak full width at half maximum [FWHM, see Fig.~\ref{fig:stability}(c)]. The quantity $\Delta B=1.1~\rm{T}$ is the field range over which the ZBP is observed. We obtain $\eta\sim10$ compared to $\eta=1$ for crossing Andreev states \cite{Lee2014}. To investigate the stability of the ZBP, Figs.~\ref{fig:magneticfield}(c)-\ref{fig:magneticfield}(f) show gate scans at the marked positions in Fig.~\ref{fig:magneticfield}a. At low field ($B_\parallel > 0.4~\rm{T}$) two subgap Andreev states are present, which evolve as a function of bias and field. In Fig.~\ref{fig:magneticfield}(e), at $B_\parallel = 2.0~\rm{T}$, these states merge at zero bias over a finite gate voltage range, distinct from the simple pointlike crossing in Fig.~\ref{fig:magneticfield}(d). Further increasing the field [$2.2~\rm{T}$ in Fig.~\ref{fig:magneticfield}(f)] has a negligible effect on the ZBP, with only the bounding gap shrinking slightly. Using a gate lever arm of $0.022$ obtained from the slope of the discrete states in Fig.~\ref{fig:magneticfield}(c), the gate voltage extent of the ZBP ($3~\rm{mV}$) can be converted in an energy range of $66~\rm{\mu eV}$. This value gives an estimate for the helical gap opening in the band structure of the wire.

To further investigate the origin of the ZBP, we vary the magnetic field orientation $\theta$ in the 2DEG plane, with $\theta=0$ being parallel to the wire. As explained above, a MZM should only manifest itself for a sufficiently strong field along $\theta=0$. Figure~\ref{fig:stability} shows three such rotations for constant magnetic field amplitudes $B_{r}$. In all cases, the rotations demonstrate the ZBP stability within a narrow angle range centered at $\theta=0$, expanding with $B_{r}$, consistent with a larger field component perpendicular to ${\bf B}_{\rm SO}$ \footnote{The finite angular offset ($\sim-4^\circ$) observed is likely due to experimental uncertainty in determining the relative orientation between the magnet and wire axes.}. For larger misalignment angles, the superconducting gap softens and the ZBP splits into two Andreev levels.

Similar to previous observations \cite{Mourik2012,Deng2016}, the height of the ZBP is significantly reduced from the quantized value of $2e^2/h$ predicted in the absence of disorder at zero temperature \cite{Law2009,Flensberg2010}. Disorder in the present samples appears comparable to conventional nanowires, as suggested by the observation of clear conductance plateaus and a hard superconducting gap. Despite this, the limited gate voltage range over which the ZBP appears is indicative of significant subband mixing. Figure~\ref{fig:stability}(c) (top panel) shows the evolution of the ZBP of Fig.~\ref{fig:magneticfield}(a) for $B_\parallel = 2.0~\rm{T}$ as a function of mixing chamber temperature $T_{\rm{mc}}$, with line cuts shown in Fig.~\ref{fig:stability}(b). The ZBP is fully suppressed by $300~\rm{mK}$ while the superconducting gap persists up to $1~\rm{K}$, with an overall lifting of the gap background due to thermal quasiparticle excitation. Figure~\ref{fig:stability}(c) (bottom panel) shows the peak height $G_{\rm{max}}$ and FWHM for $T\leq200~\rm{mK}$, where the quasiparticle background conductance is negligible. Decreasing the temperature, the ZBP gets sharper and its height monotonically increases, with a saturation reached below $50~\rm{mK}$. In this intermediate regime, the peak conductance is roughly proportional to $T^{-\alpha}$ with $\alpha\sim0.4$ while the peak FWHM scales as $G_{\rm{max}}^{-1}$, indicative of weak coupling to the lead.

In conclusion, we investigated the emergence of a ZBP from coalescing Andreev states in devices defined by top-down lithographic patterning of hybrid InAs/Al two-dimensional heterostructures. The behavior is consistent with the emergence of MZMs, with a nonuniversal conductance peak height, as seen in previous studies in individual nanowires. A top-down fabrication approach opens the door to complex device geometries and extended networks of topological devices.

\begin{acknowledgments}
We acknowledge fruitful discussions with Karsten Flensberg and Michael Hell. Research supported by Microsoft Project Q and the Danish National Research Foundation. C.M.M. acknowledges support from the Villum Foundation. F.N. acknowledges support from a Marie Curie Fellowship (Grant No. 659653).
\end{acknowledgments}

\bibliography{bibliography}

\clearpage
\onecolumngrid
\begin{center}
\textbf{\large Supplementary Material: Zero-Energy Modes from Coalescing Andreev States \\ in a Two-Dimensional Semiconductor-Superconductor Hybrid Platform}
\end{center}
\appendix

\setcounter{equation}{0}
\setcounter{figure}{0}
\setcounter{table}{0}
\makeatletter
\renewcommand{\theequation}{S\arabic{equation}}
\renewcommand{\thefigure}{S\arabic{figure}}

\onecolumngrid
\section{Wafer stack}
The wafer structure used for this work was grown by molecular beam epitaxy and consists (from top to bottom) of 10~nm Al, 10~nm $\rm{In_{0.81}Ga_{0.19}As}$, 7~nm InAs (quantum well), 4~nm $\rm{In_{0.81}Ga_{0.19}As}$, and an InAlAs buffer on an InP substrate \cite{Shabani2015,Kjaergaard2016,Kjaergaard2016b}. We stress that the top Al layer is grown directly in the growth chamber, without breaking vacuum. The electron mobility, measured in a gated Hall bar geometry with the Al removed, peaked at $\mu=20.000~\rm{cm^2 V^{-1}s^{-1}}$ for an electron density $n=9.5\times 10^{11}~\rm{cm^{-2}}$. Spin-orbit coupling was characterized in a similar wafer stack \cite{Shabani2015} by weak anti-localization measurements, obtaining a spin-orbit length $l_{\rm{SO}}=45~\rm{nm}$, corresponsing to a Rashba parameter $\alpha=42~\rm{meV nm}$.
\newline

\section{Sample preparation}
Utilizing conventional electron beam lithography techniques, mesas were patterned and etched using a typical III-V wet etchant (220:55:3:3 $\rm{H_2O:C_6H_8O_7:H_3PO_4:H_2O_2}$). Subsequently an etch mask was defined and Al was etched using a selective Al etchant (Transene-D) at $50^\circ\rm{C}$. The devices were then covered with a 40~nm layer of Al$_2$O$_3$ grown at $90^\circ\rm{C}$ using atomic layer deposition. Finally Ti/Au ($5/200~\rm{nm}$) gates were defined and deposited using electron beam evaporation. Four lithographically similar devices were used for this study, of which two showed stable ZBPs in an in-plane field, reproducibly over two cooldowns. Data on the additional devices are presented in this Supplementary Material.
\newline

\section{Electrical measurements}
Measurements were performed using standard DC and low-frequency ($f<100~\rm{Hz}$) lock-in techniques in a dilution refrigerator with a base temperature $\sim30~\rm{mK}$. An AC source drain bias of $5~\rm{\mu V}$, superimposed on a DC voltage $V_{\rm{SD}}$, was applied across the sample while the AC current $I_{\rm{SD}}$ flowing in the sample and the AC four terminal voltage $V_{\rm{4T}}$ were recorded. Conductance measurements shown throughout refer to the quantity $G=\partial I_{\rm{SD}}/ \partial V_{\rm{4T}}$.
The magnetic field was controlled using a three axis vector magnet providing a magnetic field up to $6~\rm{T}$ along the wire direction and $1~\rm{T}$ in the plane perpendicular to the wire. Therefore, field rotations as those shown in Fig.~4a of the Main Text could only be performed in a limited angle range centered around $\theta=0$. 

\begin{figure*}[b!]
\includegraphics{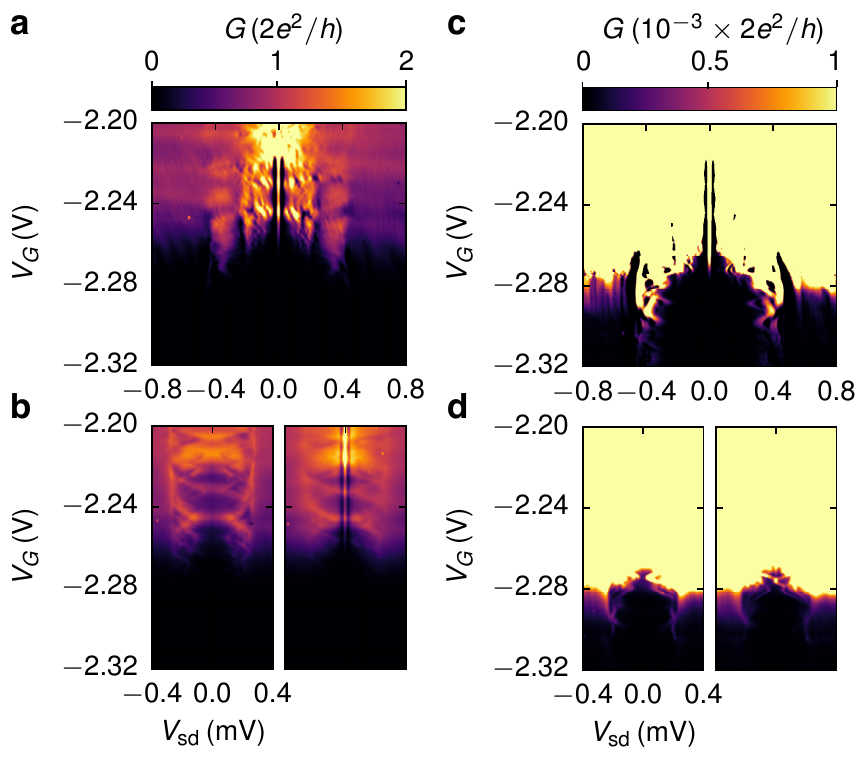}
\caption{(a),(b) The same data as in Fig.~2(b) and (c) respectively of the Main Text but with a linear color scale. (c),(d) The same as in (a),(b), but with a reduced color scale range, focusing on the low conductance regime.}
\label{fig:fig2linscale}
\end{figure*}

\begin{figure*}[b!]
\includegraphics{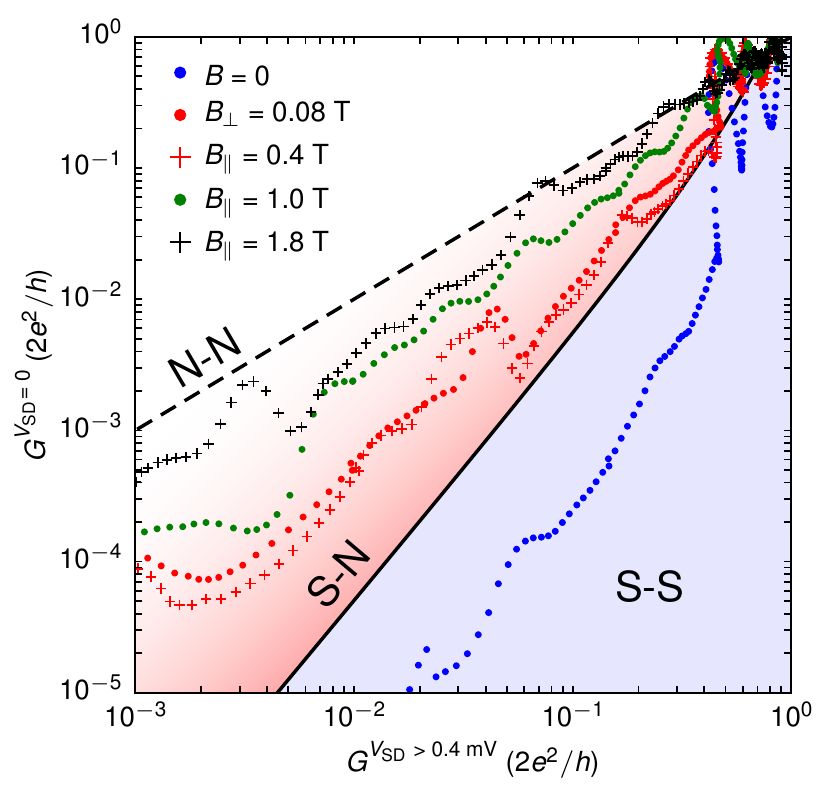}
\caption{Zero bias conductance $G^{V_{\rm SD}=0}$ as a function of normal state conductance $G^{V_{\rm SD}>0.4~{\rm mV}}$ in various field configurations. The shaded blue region denotes the expectation for a superconductor-superconductor ($S-S$) junction. The solid black line is the expectation for a single mode superconductor-normal ($S-N$) junction \cite{Beenakker1992}. The dashed black line is $G^{V_{\rm SD}=0}$ = $G^{V_{\rm SD}>0.4~{\rm mV}}$ as expected for a normal-normal junction ($N-N$).}
\label{fig:hardnesssupp}
\end{figure*}

\section{Single channel junction}
As shown in Fig.~1(c) of the Main Text, the geometry of our sample allows for the formation of a single mode quasi-ballistic junction. For completeness, we reproduce in Fig.~\ref{fig:fig2linscale} the same data as in Figs.~2(b),(c) of the Main Text but with linear color scales.

In case of a junction connecting two normal metals, it is well known that the conductance $G_{\rm{N}}$ is proportional to the junction transmission $T$. This is not the case for a junction connecting a normal metal to a superconductor. In this scenario, the conductance $G_{\rm{S}}$ is linked to the normal state conductance $G_{\rm{N}}$ by \cite{Beenakker1992}: 
\begin{align}
G_{\rm{S}} = 2G_0\frac{(G_{\rm{N}})^2}{(2G_0 - G_{\rm{N}})^2}
\label{eq:sqpcn}
\end{align}
where $G_0 = 2e^2/h$. In our experiments we associate $G_{\rm{S}}$ with the zero bias conductance ($G^{V_{\rm SD}=0}$) and $G_{\rm{N}}$ with the conductance measured at source drain biases larger than the superconducting gap ($G^{V_{\rm SD}>\Delta}$).
Figure~\ref{fig:hardnesssupp} shows a parametric plot of $G^{V_{\rm SD}>\Delta}$ versus $G^{V_{\rm SD}=0}$ for various magnetic field configurations studied in the Main Text, together with the expectation of Eq.~\ref{eq:sqpcn} (solid black line).

In the $S-S$ configuration (blue dots), $G^{V_{\rm SD}=0}$ largely increases for high transmission due to the presence of a supercurrent (not shown in Fig.~\ref{fig:hardnesssupp}). On the other hand, for low transmission, the conductance in the S-S configuration is suppressed below the S-N expectation owing to the gapped densities of states on either side of the junction. The regimes attributed in the Main Text to $S-N$ and $S-N^*$ behavior, $B_\perp=0.08~{\rm T}$ (red dots) and $B_\parallel=0.4~{\rm T}$ (red pluses) respectively, are both in good agreement with the theoretical expectation for a single mode $S-N$ junction over two orders of magnitude. For larger in-plane fields (green dots for $B_{\parallel}=1.0~\rm{T}$ and black pluses for $B_{\parallel}=1.8~\rm{T}$), relevant for accessing the topological regime, the superconducting gap softens and the in-gap conductance behaves similarly to the S-N case. This allows us to perform direct tunneling spectroscopy and observe Majorana modes at $V_{\rm{SD}}=0$. The softening of the gap for large in-plane magnetic field is consistent with recent experiments on quasi-ballstic nanowire junctions \cite{Zhang2016}. As a guide to the eye, we also plot the proportional relation expected for a N-N junction (dashed black).

\begin{figure}[tb!]
\includegraphics{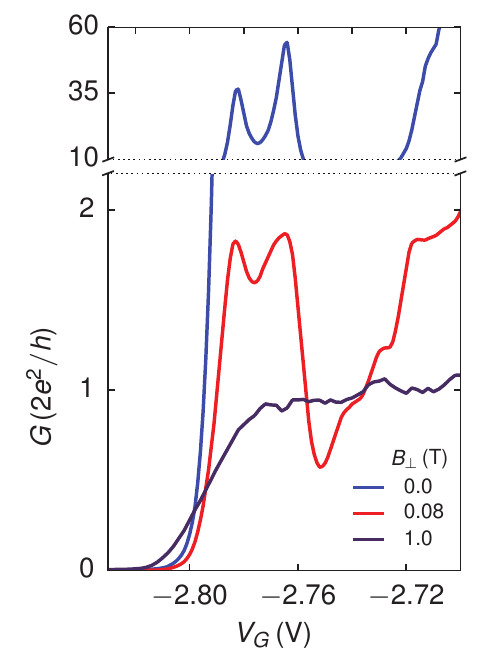}
\caption{Conductance as a function of gate voltage measured on device D2 for $B=0$ (blue), $B_\perp=0.08~\rm{T}$ (red) and $B_\perp=1~\rm{T}$ (purple).}
\label{fig:otherqpc}
\end{figure}

Similar data for another device D2, showing the zero bias conductance as a function of gate voltage for various perpendicular fields is shown in Fig.~\ref{fig:otherqpc} (cf. Fig. 1).

\section{Superconducting transitions}
\begin{figure*}[tb!]
\includegraphics{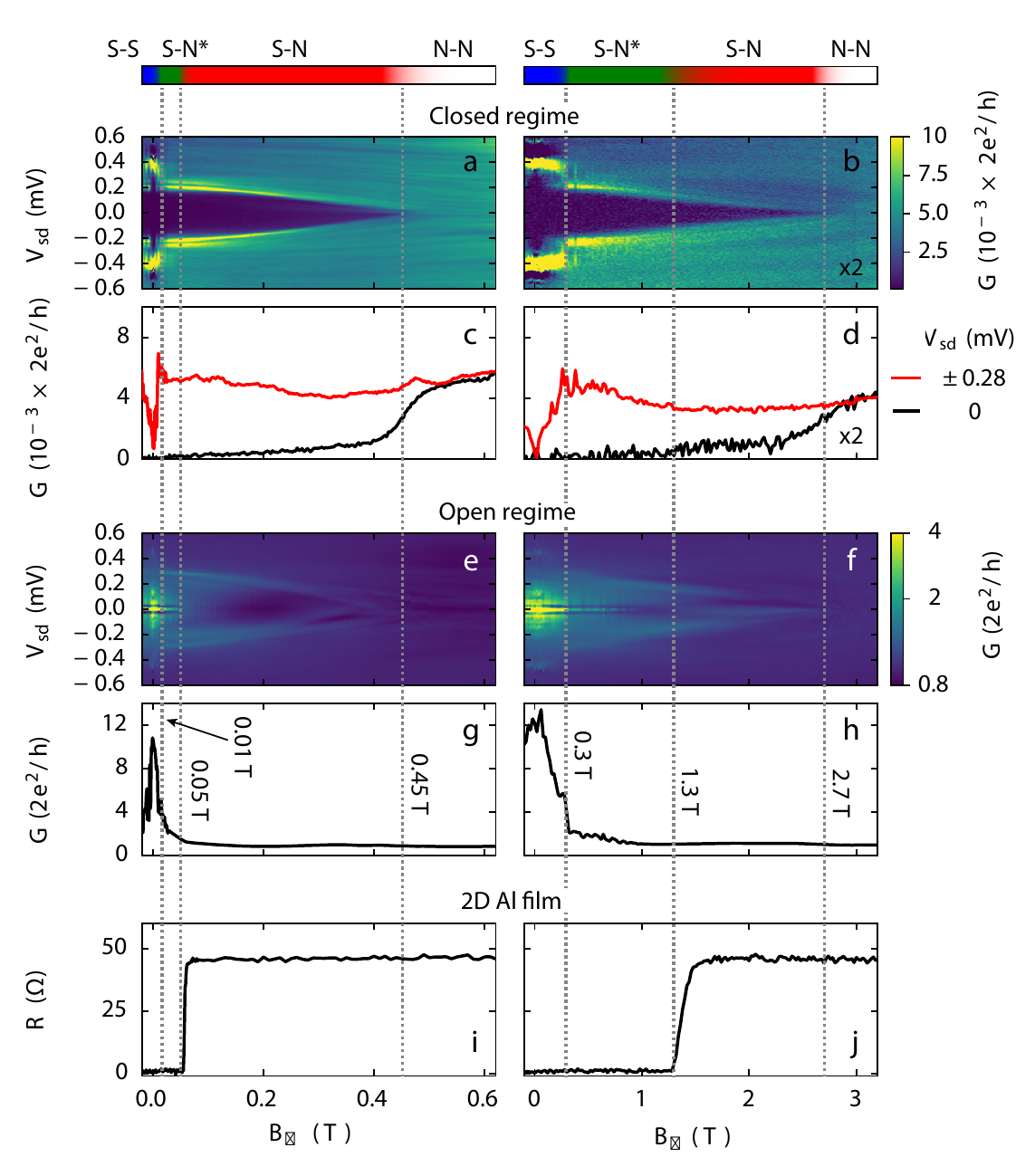}
\caption{Conductance as a function of bias and perpendicular (a) and parallel (b) magnetic fields in the tunneling regime. Line cuts are shown in (c) and (d), at both $V_{\rm sd} = 0$ (black line) and $V_{\rm sd}=\pm0.25~{\rm mV}$ (red). Note that in (b) and (d) the conductance has been scaled up by a factor of two to allow for plotting on the same colorscale. Similarly in (e)--(h) the dependence of the conductance on magnetic field is shown in the open regime. The resistance of the Al film is shown in (i) and (j) for $B_\perp$ and $B_\parallel$ respectively.}
\label{fig:filmcharac}
\end{figure*}

To further elucidate the mechanisms behind the magnetic field tuning of our devices, in Fig.~\ref{fig:filmcharac} we compare spectroscopic data [Figs.~\ref{fig:filmcharac}(a)-(h)] in two gate voltage regimes as a function of out-of-plane and in-plane magnetic field (left and right hand side of Fig.~\ref{fig:filmcharac}, respectively). Furthermore, we plot in Figs.~\ref{fig:filmcharac}(i),(j) the resistance of the large Al leads as a function of magnetic field, separately measured in a four terminal configuration.
Figures~\ref{fig:filmcharac}(a),(b) show spectroscopic data of the wire for very low coupling ($G^{V_{\rm SD}>0.4~{\rm mV}}\ll2e^2/h$), with line cuts at constant $V_{\rm{SD}}$ shown in Fig.~\ref{fig:filmcharac}(c),(d). In this case, the gate voltage is more negative than in Fig.~3(a) of the Main Text, and no subgap states appear. Figures~\ref{fig:filmcharac}(e),(f) and the line cuts of Figs.~\ref{fig:filmcharac}(g),(h) show results obtained for a more positive gate voltage, setting $G^{V_{\rm SD}>0.4~{\rm mV}}\approx2e^2/h$ and allowing the flow of a supercurrent (visible here as a conductance enhancement up to an order of magnitude over the normal state for $V_{\rm SD}=0$).
For perpendicular magnetic fields, the Al planes turn normal at $B_\perp=0.05~{\rm T}$. This is clearly associated to the $4\Delta$ to $2\Delta$ transition in the tunneling regime as well as the suppression of the conductance enhancement in the open regime. The gap closing and relative rise in the $V_{\rm SD}=0$ conductance for $B_\perp=0.45~{\rm T}$ marks the collapse of the superconductivity in the Al wire.
For an in-plane magnetic field, the $4\Delta$ to $2\Delta$ transition and the suppression of the supercurrent are markedly different, with only the latter coinciding with the critical field of the Al planes ($B_{\parallel}=1.3~\rm{T}$). As discussed further with reference of Fig.~\ref{fig:sqpcnands}, $B_{\parallel}=0.3~\rm{T}$ marks instead the typical field scale necessary to lift the hard gap in the superconducting density of states below the large Al planes. Above $B_{\parallel}=0.3~\rm{T}$, the wire is effectively probed by a constant density of states.

\section{S-QPC-N and S-QPC-S}
\begin{figure*}[tb!]
\includegraphics{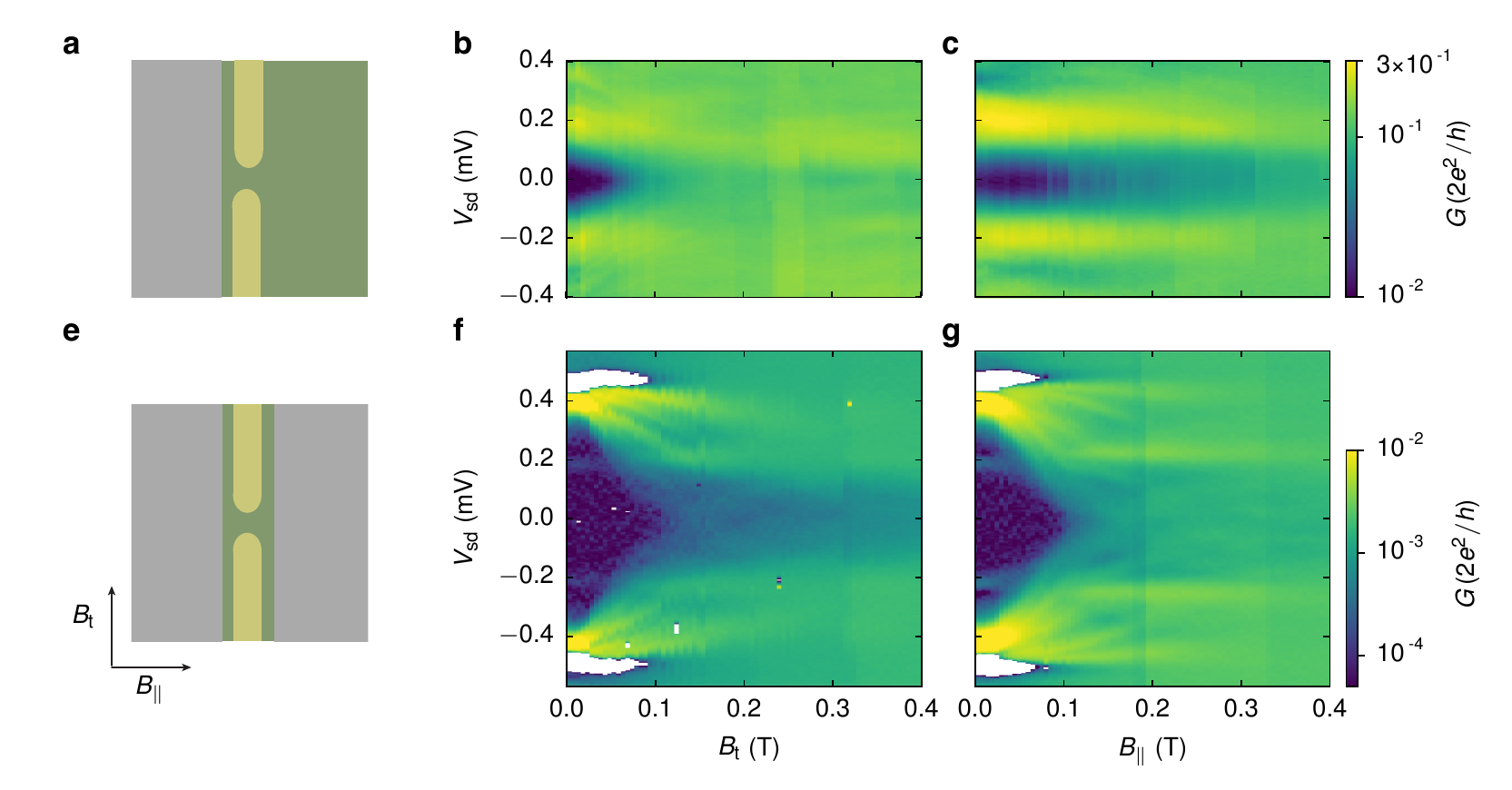}
\caption{(a) Schematic of a S-QPC-N geometry with the corresponding field dependence of the measured superconducting gap shown in (b) and (c) for magnetic fields applied along $B_{\rm t}$ and $B_\parallel$ respectively. Both fields are in the plane of the 2DEG as shown in (e). (e)--(g) similarly for a S-QPC-S geometry.}
\label{fig:sqpcnands}
\end{figure*}

To directly probe the magnetic field evolution of the superconducting density of states below a large Al plane, we perform tunneling measurements from a normal contact [Fig.~\ref{fig:sqpcnands}(a), $S-N$ configuration] and between two symmetric Al planes [Fig.~\ref{fig:sqpcnands}(e), $S-S$ configuration]. In both cases the tunneling probe is given by two evaporated Ti/Au gates defining a quantum point contact in the InAs 2DEG, similarly to Ref.~\onlinecite{Kjaergaard2016}. 
Tunneling spectroscopy as a function of an in-plane magnetic field aligned along or perpendicular to the current direction ($B_\parallel$ and $B_{\rm t}$ respectively), are shown in Figs.~\ref{fig:sqpcnands}(b),(c),(f),(g). As expected, the zero field conductance shows a $2\Delta$ gap in the S-N configuration and a $4\Delta$ in the $S-S$ configuration. Further inspection reveals the in-gap conductance suppression for the $S-S$ configuration is much stronger than in the $S-N$ case, as discussed with reference to Fig.~\ref{fig:hardnesssupp}. In both configurations, an in-plane field above $200~\rm{mT}$ lifts the in-gap conductance. In the $S-N$ configuration it is evident the inducing subgap conductance does not influence the gap size, which is largely unaffected for $B<400~\rm{mT}$. In the $S-S$ configuration, however, the convolution of the two DOS yields four peaks in conductance at $\pm 2\Delta$ and $\pm \Delta$. As the field is increased further towards $B_\parallel=0.4~{\rm T}$ the $\pm\Delta$ edges are independent of field.
In conclusion, both devices demonstrate that for magnetic fields of the order of $400~\rm{mT}$, the superconducting gap measured in a 2D geometry stays roughly constant, however with a significant increase in the subgap conductance.

\section{ZBPs in other devices}
\begin{figure*}[tb!]
\includegraphics{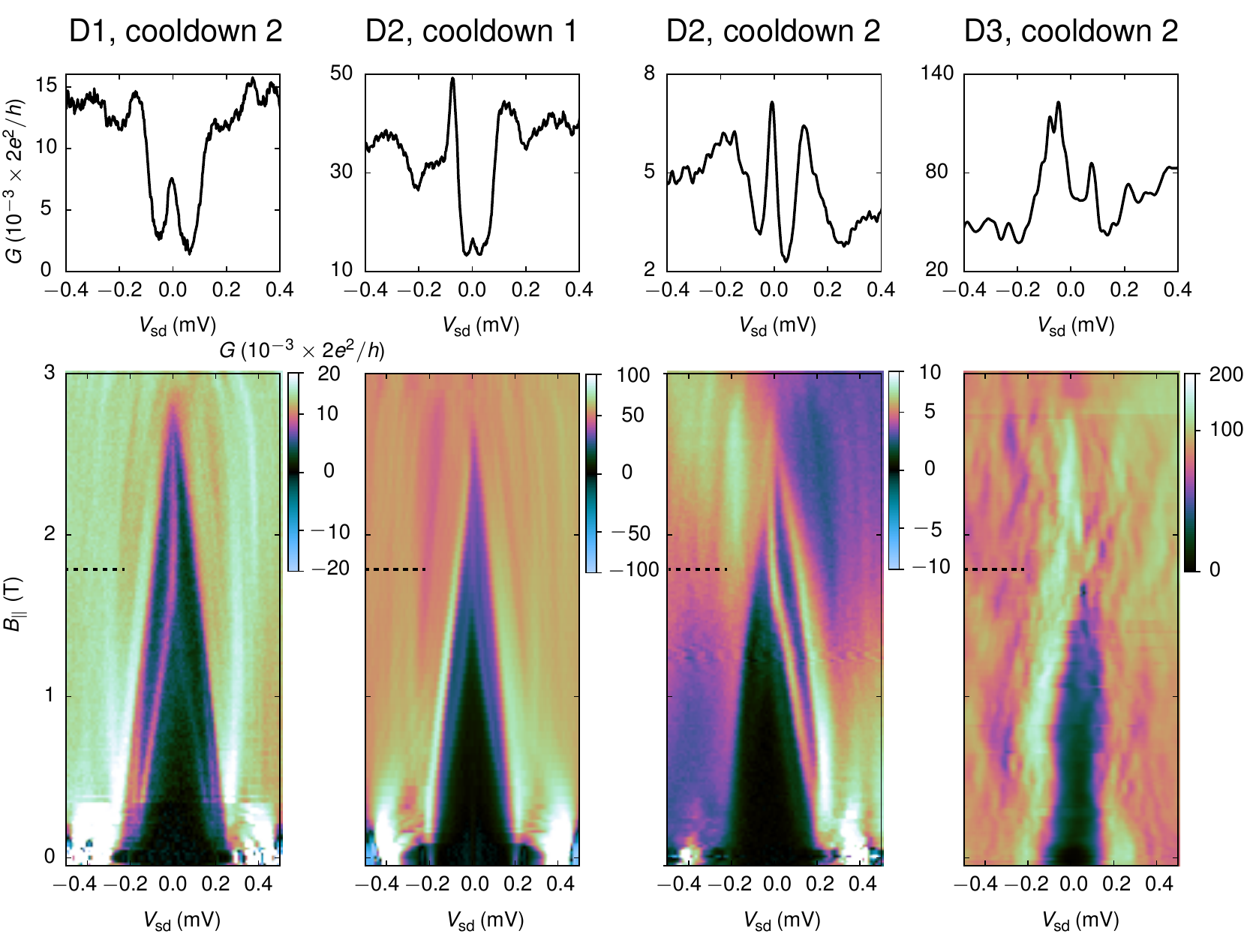}
\caption{In three separate devices, characteristic conductance line cuts as a function of bias are shown in the top row taken at $B_\parallel=1.8~{\rm T}$. The bottom row presents the full conductance maps as a function of magnetic field and bias. The dashed lines indicate $B_\parallel=1.8~{\rm T}$, the location of the linecuts in the top row. In devices D1 and D2 a ZBP is present, whilst in D3 it is absent presumably due to disorder.}
\label{fig:otherzbp}
\end{figure*}

Four devices were used for this study, named D1, D2, D3 and D4. Device D1, which showed the highest gate stability, is presented throughout the Main Text with data shown from cooldown 1. Measurements of D2 showed similar properties to the device presented in the Main Text, in two cooldowns. Devices D3 and D4 were more disordered and no stable ZBP could be identified. In Fig.~\ref{fig:otherzbp} we present conductance measurements on three devices (D1, D2, and D3) as a function of $V_{SD}$ and $B_{\parallel}$. Devices D1 and D2 demonstrate the presence of stable zero energy states in two cooldowns. Device D3 indicates the presence of a superconducting gap masked by disorder, and no ZBP can be identified, similarly to D4 (not shown).

\begin{figure*}[tb!]
\includegraphics{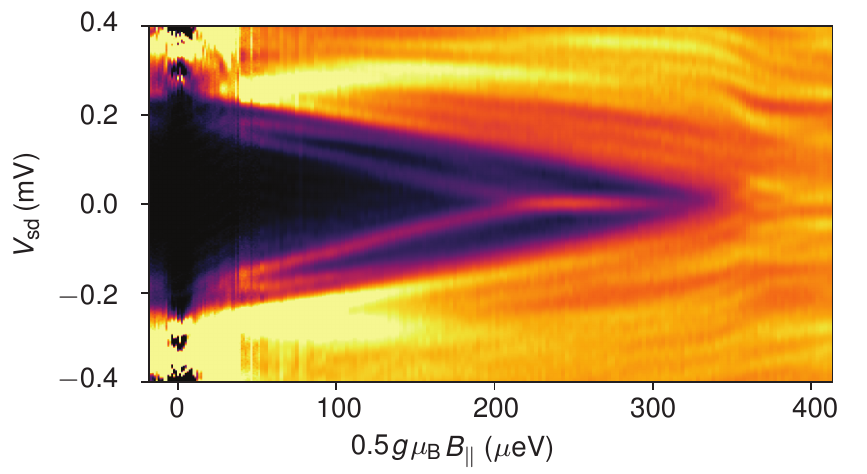}
\caption{The same data as in Fig.~3(a) of the Main Text, but with the horizontal magnetic field axis expressed in energy units. The magnetic field extension of the ZBP is several times its width.}
\label{fig:benergy}
\end{figure*}

\section{Stability of the ZBP}
In order to distinguish the ZBP presented in Fig.~3(a) of the Main Text from a zero-crossing of Andreev states, we plot the same data in Fig.~\ref{fig:benergy} with the horizontal axis converted to energy units. The extent of the ZBP as a function of energy is several times its width, indicating the observed ZBP is likely not due to a crossing of Andreev states.

\begin{figure*}[tb!]
\includegraphics{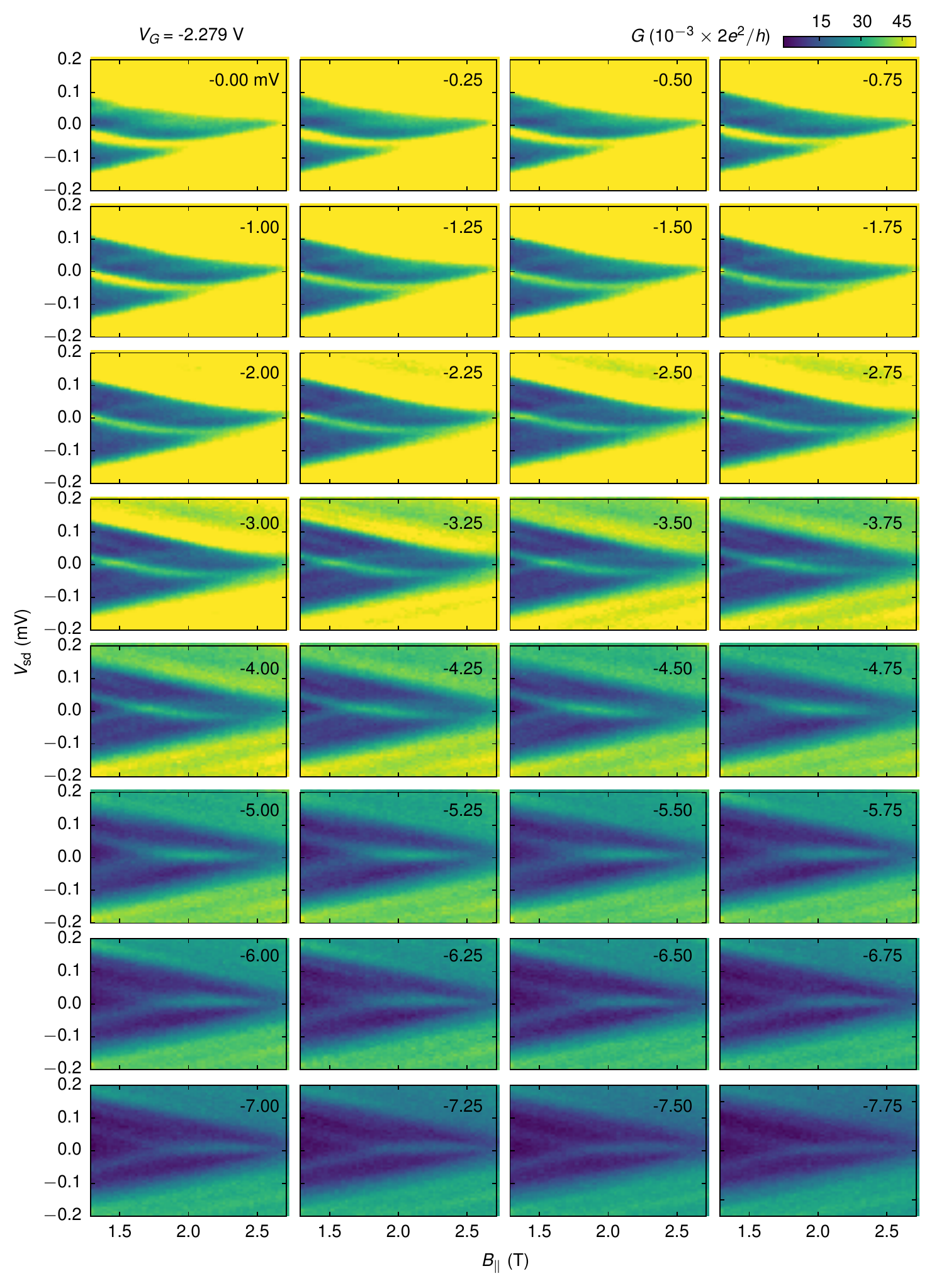}
\caption{Snapshots of the bias and magnetic field dependence for a range of gate voltages from $V_{\rm G}=-2.279~{\rm V}$ to $-2.28675~{\rm V}$. The numbers in the upper corner of each panel indicate the gate voltage offset in mV from the starting point.}
\label{fig:peakstability}
\end{figure*}

\begin{figure*}[tb!]
\centerline{\includegraphics[width=1\columnwidth]{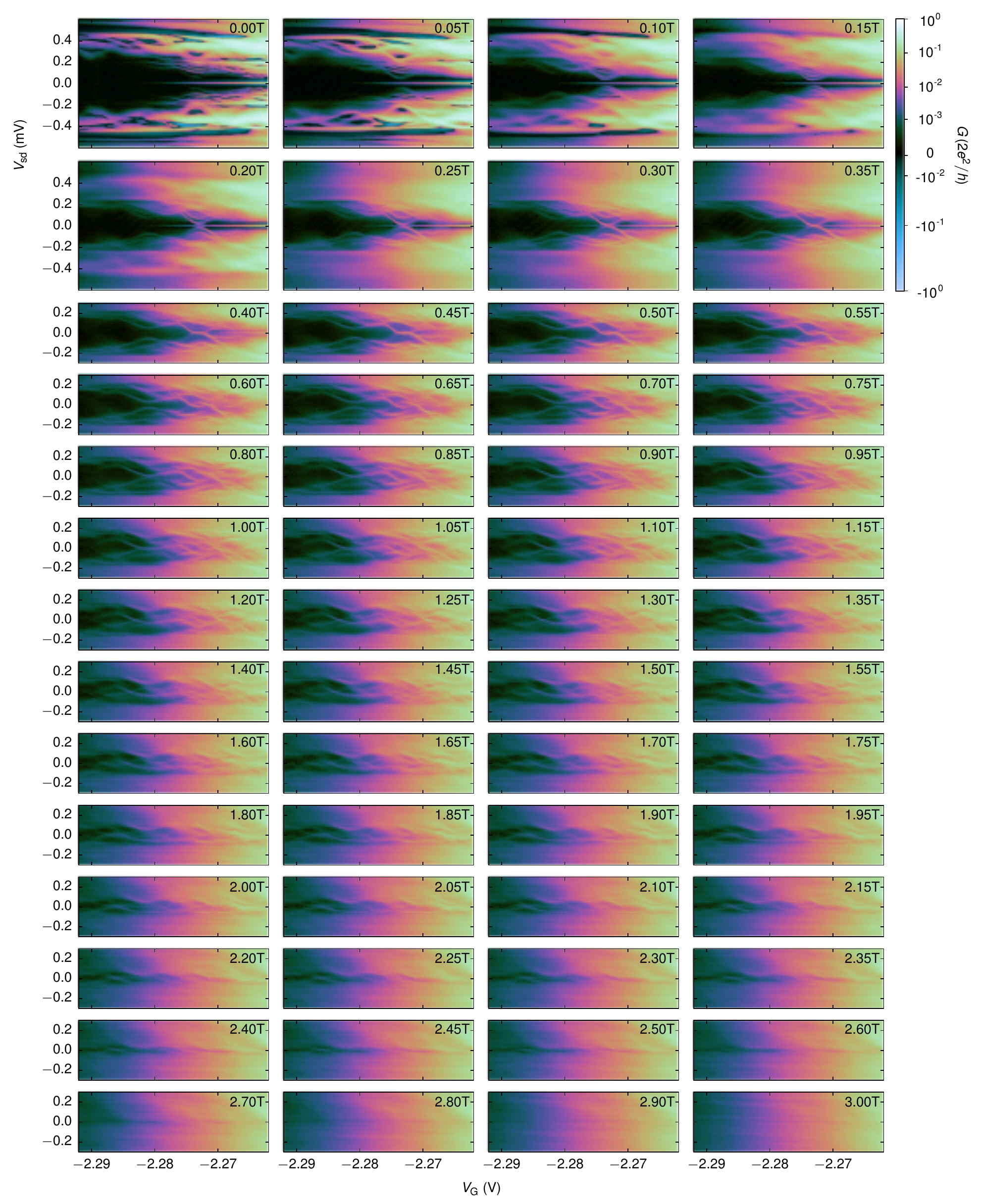}}
\caption{Snapshots of the bias and gate dependence for a range of magnetic fields indicated in the upper corner of each panel. A nonlinear colorscale is used.}
\label{fig:zbpgate}
\end{figure*}

Fig.~\ref{fig:peakstability} shows the gate and bias dependence of the ZBP discussed in the Main Text at fine gate voltage intervals, indicating its stability over a $3~\rm{mV}$ range in gate voltage. The gate voltage extent can be converted to a chemical potential span of $66~\rm{\mu eV}$ using the gate lever arm. Detailed gate dependence of the device behavior at a range of magnetic fields is presented in Fig.~\ref{fig:zbpgate}.

\end{document}